\documentclass[conference]{IEEEtran}

\usepackage{amsmath,amssymb,amsfonts}
\usepackage{algorithmic}
\usepackage{graphicx}
\usepackage{textcomp}
\usepackage{xcolor}

\usepackage[utf8]{inputenc}
\usepackage[T1]{fontenc}
\usepackage{hyperref}
\hypersetup{hidelinks,unicode,breaklinks}
\usepackage{acronym}
\usepackage{dblfloatfix}
\usepackage{tabto}
\usepackage{enumerate}
\usepackage{siunitx}
\usepackage{orcidlink}

\usepackage{tikz}
\usetikzlibrary{math,arrows.meta,calc} 

\usepackage{colortbl} 

\usepackage{flushend} 

\usepackage{cleveref}

\acrodef{AES}{Advanced Encryption Standard}
\acrodef{CBC}{Cipher Block Chaining}
\acrodef{CTR}{Counter}
\acrodef{ECB}{Electronic Codebook}
\acrodef{GCM}{Galois/Counter Mode}

\newcommand{\at}{\makeatletter@\makeatother}

\DeclareMathOperator{\iv}{iv}
\DeclareMathOperator{\key}{k}
\DeclareMathOperator{\enc}{Enc}

\crefname{subsection}{subsection}{subsections}
\Crefname{subsection}{Subsection}{Subsections}

\usepackage[style=ieee, backend=biber, url=false, hyperref, maxnames=5, minnames=1, maxcitenames=2, mincitenames=1, defernumbers=true]{biblatex}
\makeatletter
\renewcommand{\fnum@figure}{Figure~\thefigure}
\makeatother

\usepackage{ifpdf}
\ifpdf
\pdfoutput=1 
\pdftrue
\fi

\makeatletter
\def\ps@IEEEtitlepagestyle{
  \def\@oddfoot{\mycopyrightnotice}
  \def\@evenfoot{}
}
\def\mycopyrightnotice{
  {\footnotesize
    \begin{minipage}{0.8\textwidth}
    \centering
    Please cite as: \fullcite{selfref}.
    \end{minipage}
  }
}
\makeatother
\makeatletter
\let\blx@rerun@biber\relax
\makeatother
\usepackage[shortcuts]{extdash} 

\DeclareBibliographyCategory{selfref}
\addtocategory{selfref}{selfref}

\title{\bfseries\Large Distinguishing Tor From Other Encrypted Network Traffic\\Through Character Analysis}
\author{%
    \large Pitpimon Choorod$^1$\,\orcidlink{0000-0002-9279-0710}, Tobias J. Bauer$^2$\,\orcidlink{0009-0006-1073-3971}, and Andreas Aßmuth$^3$\,\orcidlink{0009-0002-2081-2455}\\[0.3ex]\normalsize\normalfont
        $^1$King Mongkut's University of Technology North Bangkok, Prachinburi, Thailand\\
        Email: {\tt pitpimon.c@itm.kmutnb.ac.th}\\
        $^2$Fraunhofer Institute for Applied and Integrated Security, Weiden, Germany\\
        Email: {\tt tobias.bauer@aisec.fraunhofer.de}\\
	$^3$Ostbayerische Technische Hochschule Amberg-Weiden, Amberg, Germany\\
	Email: {\tt a.assmuth@oth-aw.de}%
}

\begin{document}

\maketitle

\begin{abstract}
For journalists reporting from a totalitarian regime, whistleblowers and resistance fighters, the anonymous use of cloud services on the Internet can be vital for survival. The Tor network provides a free and widely used anonymization service for everyone. However, there are different approaches to distinguishing Tor from non-Tor encrypted network traffic, most recently only due to the (relative) frequencies of hex digits in a single encrypted payload packet. While conventional data traffic is usually encrypted once, but at least three times in the case of Tor due to the structure and principle of the Tor network, we have examined to what extent the number of encryptions contributes to being able to distinguish Tor from non-Tor encrypted data traffic.
\end{abstract}

\renewcommand\IEEEkeywordsname{Keywords}

\section{Introduction}\label{sec:introduction}
When it comes to security for cloud services, most people first think of ensuring the security goals of confidentiality, integrity, availability, and authenticity. However, anonymization services also play an important role, as they are used, for example, by journalists or opponents of regimes in authoritarian countries to access cloud services and to provide information about grievances in the country in question. In general, anonymization services can increase the privacy of users of cloud services, as anonymization prevents unauthorised third parties from tracking or profiling users based on their cloud-related activities. To be fair, it must also be noted at this point that criminals also have an interest in anonymization services, whether to conceal their criminal activities or to set up and operate largely anonymous trading platforms on the Darknet, e.g., Silkroad~\cite{Liggett2020}.\par
The Tor project (Tor, short for The Onion Router) has been a very popular and free anonymization service on the Internet for years. The basic idea of anonymization can be described as follows: a number of $n$ nodes of the Tor network are identified via which communication is to take place, for example accessing a website via http. Depending on the number $n$ (the default is $n=3$), the actual request is encrypted $n$ times in succession, creating $n$ (encryption) layers~-- like an onion. Each node of the identified path through the Tor network now removes one of these layers by decrypting it before forwarding the data to the next Tor node. The last layer is finally removed by the exit node, whose IP address is then visible when accessing the actual website, but not the IP address of the actual user's computer. Each node in the Tor network only knows its predecessor and its successor for the respective path. For a detailed description of how Tor works, please refer to \cite{Dingledine2004} and, of course, to the documentation of the Tor project~\cite{TorProject}.\par 
Against the background described above, the question can now be asked whether and how it is possible to distinguish Tor traffic from otherwise encrypted traffic when monitoring network traffic. This question has a fundamental core aspect for cryptography. The definition of perfect secrecy goes back to Shannon~\cite{Shannon1949}. The necessary and sufficient condition for perfect secrecy is $ \operatorname{Pr}(C=c\,|\,M=m) = \operatorname{Pr}(C=c) $, where $\operatorname{Pr}(C=c)$ is the (a priori) probability of obtaining the ciphertext $c$, and $\operatorname{Pr}(C=c\,|\,M=m)$ is the conditional probability of ciphertext $c$ if message $m$ was chosen for encryption. Building on this theorem, modern textbooks on cryptography describe experiments that are the basis for security definitions for encryption schemes as we use them today. As an example of such an experiment, the so-called adversarial indistinguishability experiment for probabilistic symmetric-key encryption schemes is presented here (according to~\cite{KatzLindell2020}):
\begin{enumerate}
    \item An attacker $\mathcal{A}$ chooses two messages $m_0$ and $m_1$ of the same length for a given encryption scheme with security parameter $N$. The security parameter may be viewed as corresponding to the length of the key.
    \item A random key $k$ is generated (depending on $N$) and a bit $b\in\lbrace0,1\rbrace$ is chosen at random. $\mathcal{A}$ receives the so-called challenge ciphertext $c\leftarrow\operatorname{Enc}_k(m_b)$.
    \item $\mathcal{A}$ outputs a bit $b'\in\lbrace0,1\rbrace$.
    \item The result of the experiment is $1$ if $b=b'$, otherwise $0$.
\end{enumerate}
In the case of perfect secrecy according to Shannon's theorem, the result of the experiment corresponds to the guess probability of \qty{50}{\percent}. For security definitions for modern encryption schemes, the probability for result $1$ must be increased slightly, whereby this increase is set via the security parameter $N$ and its actual value is a negligible function in $N$ for all realistic adversaries. Now imagine running this experiment in parallel for two different encryption schemes, where in 1) the length of all messages is the same, and the result of both experiments is only slightly more than \qty{50}{\percent} for all realistic adversaries~-- if $\mathcal{A}$ cannot practically decide which of the messages led to $c$ by encryption, can $\mathcal{A}$ distinguish which challenge ciphertext was generated by which encryption scheme? According to Rogaway, several modes of operations for modern blockciphers achieve computational indistinguishability from random bits~\cite{Rogaway2011}. So, if $\mathcal{A}$ cannot distinguish any ciphertext from a random bitstring of the same length, it should not be feasible to distinguish Tor-encrypted network traffic from otherwise (non-Tor) encrypted network traffic.\par 
The paper is structured in the following manner: \Cref{sec:relatedwork} provides an overview of published work that deals with the distinction between Tor and non-Tor encrypted traffic. \Cref{sec:preliminarywork} summarises the most important results of a novel approach for classifying Tor and non-Tor traffic presented by Pitpimon Choorod in her PhD thesis~\cite{Choorod2023}. Based on these results, new experiments have been carried out which are presented in \Cref{sec:experiments}. Finally, \Cref{sec:conclusion} ends with a conclusion and an outlook on future work.

\section{Related Work}\label{sec:relatedwork}
The robustness of encryption schemes has led researchers to study the Tor traffic classification domain using flow-based or packet-based features. Lashkari et al.~\cite{lashkari2017characterization}, the creators of the University of New Brunswick, Canadian Institute for Cybersecurity (UNB-CIC) dataset, achieved high performance in detecting Tor traffic using time-based features and attained precision and recall rates above $0.9$ with the C4.5 algorithm. Using the same dataset, Kim et al.~\cite{kim2018tor} instead focused on payload-based features with the first $54\,\text{bytes}$ of TCP packet headers as input. The results indicated that the one-dimensional convolutional neural network model outperformed the C4.5 algorithm of \cite{lashkari2017characterization}, achieving precision and recall rates of $1.0$ in classifying both Tor and non-Tor traffic. Hu et al.~\cite{hu2020} expanded the scope of Darknet traffic analysis. They distinguished four Darknet traffic types including Tor, I2P, ZeroNet, and Freenet using 26 time-based flow features, achieving an accuracy of \qty{96.9}{\percent}. However, while flow features are effective in classifying Tor traffic, factors such as network sensitivities, including asymmetric routing, can undermine the reliability of time-based features. The approach chosen in~\cite{Choorod2023} addresses this limitation by enhancing reliability under diverse network conditions. We focus on extracting non-timing related features from the encrypted data within packet payloads, thereby presenting a challenge to the conventional assumptions of Shannon's theorem.

\section{Preliminary Work}\label{sec:preliminarywork}
This section describes the preliminary work that Pitpimon Choorod carried out as part of her PhD thesis~\cite{Choorod2023}. A publication summarising the key aspects of the PhD thesis is also available~\cite{choorod2024classifying}.\par 
In computer networks, data payloads are commonly represented as hexadecimal characters, using a base-16 numbering system that ranges from \textit{0} to \textit{9} and \textit{a} to \textit{f}. The study focused on analysing these hexadecimal characters in their single-digit (1-hex) form extracted from encrypted data. To facilitate the analysis, two key statistical features were used: 1) a frequency set feature, which consists of 16 individual features for quantifying the occurrence of each hexadecimal character within data payloads, and 2) a frequency ratio set, also including 16 features, for calculating the proportion of each character’s frequency relative to the total character count within a payload. The normalisation of these frequencies was crucial to ensure length normalisation, thereby minimising potential biases in analysing encrypted payloads that could arise from relying solely on their absolute packet lengths.

The analysis utilised two data sources to validate the robustness and reliability of the results. The first was a public Tor dataset from the UNB-CIC, where network traffic was categorised into eight application types (audio, browsing, chat, email, FTP, P2P, video, and VoIP). In addition to the public dataset, a private dataset was created by capturing Tor-encrypted traffic data packets using Wireshark. The corresponding data consists of browsing applications. \Cref{instance} presents the number of instances for the eight application types in the public dataset and one application type in the private dataset. It should be noted that the instances for both Tor and non-Tor are balanced.

\begin{table}[h]
\centering
\caption{Number of balanced Tor and non-Tor instances\\for nine applications}
\label{instance}
\begin{tabular}{|l|l|l|l|l|l|}
\hline
\cellcolor[HTML]{EFEFEF} Audio & 26,082 & \cellcolor[HTML]{EFEFEF} Email & 12,300 & \cellcolor[HTML]{EFEFEF} Video & 32,154 \\ \hline
\cellcolor[HTML]{EFEFEF} Browsing & 71,950 & \cellcolor[HTML]{EFEFEF} FTP & 514,952 & \cellcolor[HTML]{EFEFEF} VoIP & 737,382 \\ \hline
\cellcolor[HTML]{EFEFEF} Chat & 6,504 & \cellcolor[HTML]{EFEFEF} P2P & 433,770 & \cellcolor[HTML]{EFEFEF} Private & 29,600 \\ \hline
\end{tabular}

\end{table}

According to \Cref{sec:introduction}, the investigation commenced with the assumption that there is no difference between Tor and non-Tor traffic in terms of encrypted payloads. Initially, descriptive statistics were utilised to describe and summarise the characteristics of the sample data. In this study, statistical measurements including character distribution were employed, which helps reveal patterns in how individual features are spread across the range of values in both Tor and non-Tor encrypted payloads. The mean measurement indicates the central tendency of each feature, aiding in identifying the average value of the individual features. Standard deviation measures the variability or dispersion of each feature, highlighting trends or patterns. Minimum and maximum values provide insights into the range of features within the Tor and non-Tor datasets, with a wider range potentially indicating greater variability in traffic characteristics. The results clearly showed that all measurements of Tor and non-Tor were significantly different, except for the ratio features. This exception can be attributed to the effect of normalisation, which tends to minimise discrepancies in data scale and distribution, thereby making the ratio features appear more similar across both datasets. Additionally, these findings were generalised using the Mann-Whitney test, which revealed a significant differentiation rate between Tor and non-Tor traffic of \qty{95.42}{\percent} for the public dataset and \qty{100}{\percent} for the private dataset.

\begin{figure}[htbp]
    \centering
    \includegraphics[width=\linewidth]{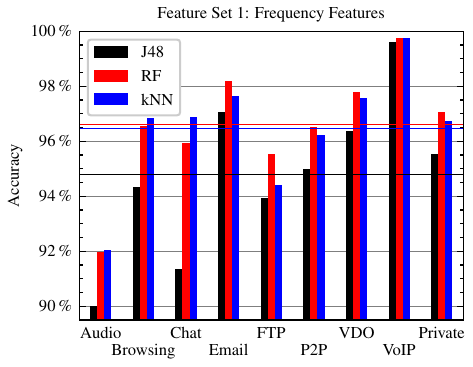}
    \includegraphics[width=\linewidth]{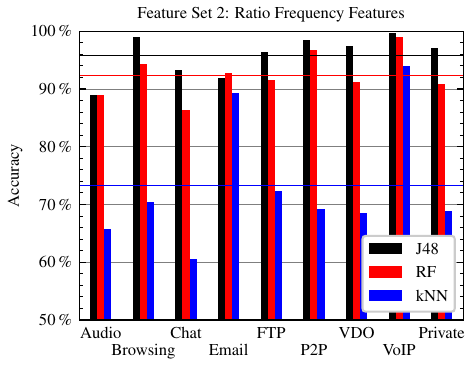}
    \vspace*{-5mm} 
    \caption{Results of the approach proposed in \cite{Choorod2023}.}
    \label{fig:ml_results}
    \vspace*{-5mm} 
\end{figure}

The second phase of the study, run in Weka~\cite{hall2009weka}, focused on classifying Tor and non-Tor traffic using machine learning. Three supervised learning algorithms were used: J48~\cite{bhargava2013decision}, Random Forest (RF)~\cite{breiman2001random}, and k-Nearest Neighbors (kNN)~\cite{guo2003knn}, with a specific focus on encrypted payload features. The J48 algorithm is identical to the aforementioned C4.5 algorithm. As depicted in \Cref{fig:ml_results}, for the frequency feature set on the public dataset, it can be noted that classification accuracy exceeded \qty{90}{\percent} for all models across all applications. Notably, both RF and kNN achieved the highest score of \qty{99.77}{\percent} for VoIP. For the private dataset, RF demonstrated superior performance with a score of \qty{97.06}{\percent}.

Regarding the ratio frequency feature set on public datasets, all models surpassed a classification accuracy of \qty{60.62}{\percent} across all applications, with J48 achieving the best score of \qty{99.63}{\percent} for VoIP. For the private dataset, J48 reached an impressive accuracy of \qty{97.12}{\percent}. These results show a similar trend in both public and private datasets, ensuring the consistency of these findings.

These results conclusively demonstrate that Tor and non-Tor traffic are statistically distinct, enabling efficient classification of both types of traffic using features derived exclusively from a single encrypted payload packet.

\section{New Experiments}\label{sec:experiments}
One might be tempted to explain these results by the fact that encryption in Tor and non-Tor encrypted traffic in practice does not show the desired property presented in \Cref{sec:introduction}. If we compare Tor to non-Tor encrypted traffic, the main difference is that while, e.g., TLS traffic is encrypted only once Tor traffic is encrypted multiple times because of the onion-like layer model. Therefore, in this paper we focus solely on distinguishing between single-encrypted data and triple-encrypted data. We demonstrate that data that was encrypted one time has the same statistical properties as data that was encrypted three times. This results in machine learning algorithms being unable to distinguish between them. We have tested the \ac{AES} algorithm in various modes of operation.

As a first step, sample data needed to be generated. In order to study the effect of the underlying data, our generation method outputs two sets of data samples. Each set contains $\# = 10^6$ strings of $l = 512$~bytes. The length $l$ was chosen to be a multiple of the \ac{AES} block size of $128\,\text{bits}$ and following the message length in the Tor specification~\cite{TorSpec}. The first set is generated using the cryptographically secure pseudorandom number generator \textit{/dev/urandom} of Linux, whereas the second set contains the same amount of samples, but each sample is a string of null-bytes, representing data with zero randomness to it. We denote a sample of the first set as $r_i^0$ (random data) and a sample of the second set as $z_i^0$ (zeros) with $0 \le i < \#$.

Next, we generate $\#$ initialization vectors (IVs) $\iv_i^1, \iv_i^2, \iv_i^3$ and encryption keys $\key_i^1, \key_i^2, \key_i^3 \quad (0 \le i < \#)$ at random. The encryption algorithm takes any data $d$, an IV $\iv$, and an encryption key~$\key$ and outputs a ciphertext $c = \enc\left(d, \iv, \key\right)$.
We then perform a single encryption of samples $r_i^0$ and $z_i^0$ to obtain $r_i^1$ and $z_i^1$. Next, two more rounds of encryption are performed to obtain $r_i^3$ and $z_i^3$, respectively. The following equations illustrate the process:
\vspace*{-1.5mm} 
\begin{align}
    r_i^1 &= \enc\left(r_i^0, \iv_i^1, \key_i^1\right) \\
    z_i^1 &= \enc\left(z_i^0, \iv_i^1, \key_i^1\right) \\
    r_i^3 &= \enc\left(\enc\left(r_i^1, \iv_i^2, \key_i^2\right), \iv_i^3, \key_i^3\right) \\
    z_i^3 &= \enc\left(\enc\left(z_i^1, \iv_i^2, \key_i^2\right), \iv_i^3, \key_i^3\right)
\end{align}\vspace*{-5.5mm} 

We denote the sets of samples by their capital letter, i.e.,
\vspace*{-1.5mm} 
\begin{align*}
    R^1 &= \left\{r_1^1, r_2^1, \dots, r_\#^1\right\}, &R^3 &= \left\{r_1^3, r_2^3, \dots, r_\#^3\right\},\\
    Z^1 &= \left\{z_1^1, z_2^1, \dots, z_\#^1\right\}, &Z^3 &= \left\{z_1^3, z_2^3, \dots, z_\#^3\right\}.    
\end{align*}
\vspace*{-6mm} 

In order to study the effect of different \ac{AES} modes of operation, we perform these preparatory steps for each mode. In total, we opted to study these three modes: \ac{CBC}, \ac{CTR}, and \ac{ECB}. The \ac{CBC} mode is widely used within the TLS~1.2 specification~\cite{rfc5246} and the \ac{CTR} mode forms the basis for the \ac{GCM}~\cite{Dworkin2007}, which is used extensively throughout the Internet~\cite{F5Labs2021}. Furthermore, \ac{GCM} is used in two out of five specified cipher suites of TLS~1.3~\cite{rfc8446} and is preferred by a majority of web servers~\cite{F5Labs2021}. In addition to that, we include a third mode of operation, \ac{ECB}, in our experiments. This mode of operation is highly insecure as it leaks the equality of blocks~\cite{Rogaway2011}, does not provide the required randomization of the ciphertext, and should therefore not be used within any cryptographic protocols. However, we can show that even this mode of operation achieves the indistinguishability between single-encrypted data and triple-encrypted data provided that a random key is used.

Our experiments follow a simple four-step procedure:
\begin{enumerate}
    \item Generate a dataset with features $\mathcal{X}$ and labels $\mathcal{Y}$: \\
            $\mathcal{D} = \left\{d_1, d_2, \dots, d_{2\cdot \#}\right\} \overset{\textnormal{e.g.}}{=} R^1 \cup R^3$ \\
            $\mathcal{X} = \left\{X_1, X_2, \dots, X_{2 \cdot \#}\right\},\quad X_i = F\left(d_i\right)$ \\
            $\mathcal{Y} = \left\{y_1, \dots, y_{2\cdot\#}\right\} = \left\{ \begin{matrix}0 & \textnormal{~if~} d_i \textnormal{~single-encrypted}\\1 & \textnormal{~if~} d_i \textnormal{~triple-encrypted} \end{matrix} \right..$
    \item Split $\left(\mathcal{X}, \mathcal{Y}\right)$ into a training set $\left(\mathcal{X}_\textnormal{tr}, \mathcal{Y}_\textnormal{tr}\right)$ and a test set $\left(\mathcal{X}_\textnormal{te}, \mathcal{Y}_\textnormal{te}\right)$ using a 75:25-split.
    \item Fit a machine learning model to the training set.
    \item Evaluate the trained machine learning model on the test set and compute a confusion matrix.
\end{enumerate}
The function $F$ denotes the feature engineering, which is similar to the previous work~\cite{Choorod2023}. $F(d_i)$ simply counts the hexadecimal digits \textit{0}~to~\textit{f} and returns the relative frequencies for each of the 16 digits, i.e., a vector $X_i \in \mathbb{R}^{16}$. Since the original data strings are fixed-length strings of random data~$\left(r_i^0\right)$ or zeros~$\left(z_i^0\right)$, relative and absolute frequencies behave identically, which is why we used only the relative frequencies of the hexadecimal digits.

\begin{figure}[htbp]%
    \centering%
    \includegraphics[width=\linewidth]{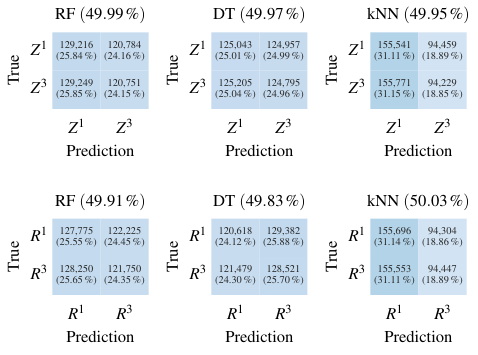}%
    \vspace*{-3mm}%
    \caption{Results with the \ac{ECB} mode of operation.}%
    \label{fig:results-ecb}%
    \vspace*{-3mm}%
\end{figure}

\begin{figure}[htbp]%
    \centering%
    \includegraphics[width=\linewidth]{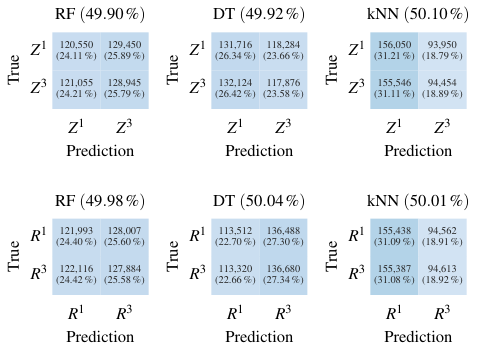}%
    \vspace*{-3mm}%
    \caption{Results with the \ac{CBC} mode of operation.}%
    \label{fig:results-cbc}%
    \vspace*{-3mm}%
\end{figure}

\begin{figure}[htbp]%
    \centering%
    \includegraphics[width=\linewidth]{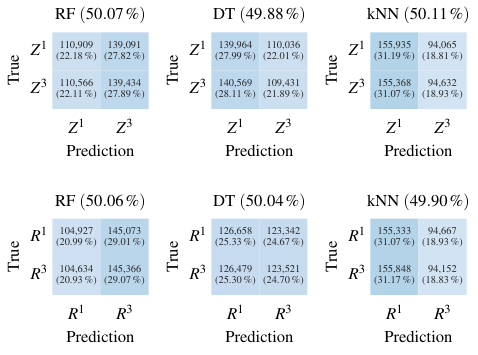}%
    \vspace*{-3mm}%
    \caption{Results with the \ac{CTR} mode of operation.}%
    \label{fig:results-ctr}%
    \vspace*{-3mm}%
\end{figure}

The results are depicted in \Cref{fig:results-ecb,,fig:results-cbc,,fig:results-ctr}. Each figure shows the results for one mode of operation using three machine learning algorithms -- Random Forest (RF), Decision Tree (DT), and k-Nearest Neighbors (kNN) -- on two datasets $\mathcal{D}_Z = Z^1 \cup Z^3$ (upper row) and $\mathcal{D}_R = R^1 \cup R^3$ (lower row). The accuracy value of each experiment is displayed in the subplot title. In order to ensure comparability with the preliminary work described in \Cref{sec:preliminarywork}, we opted to employ the same three machine learning algorithms. However, in this paper we use the \textit{scikit-learn}~\cite{scikit-learn} machine learning framework. This framework uses for Decision Tree construction the CART algorithm that is similar to C4.5 and J48, respectively~\cite{scikit-learn-dt}.

Our results show clearly that none of these machine learning models is able to distinguish between single-encrypted and triple-encrypted payload using the relative frequencies of the 16 hexadecimal digits as feature vectors. The accuracy is always about $50\,\% \pm 0.17\,\%$, which is due to run-to-run variance and does not indicate any ability to distinguish these two categories. We would like to remind the reader that $50\,\%$ is exactly the guess probability. Even the insecure \ac{ECB} mode of operation achieves the indistinguishability property described in \Cref{sec:introduction} and the machine learning models are therefore unable to predict the class correctly in significantly more than \qty{50}{\percent} of all cases (cf.~\Cref{fig:results-ecb}).\par 
These new experiments clearly indicate that the distinction between one-time and three-time encryption cannot be the decisive criterion in the generation of ciphertexts. Therefore, the reason to why the method described in \cite{Choorod2023} and abridged in \Cref{sec:preliminarywork} is able to distinguish Tor from non-Tor encrypted data traffic with such high rates must not be related to the number of encryption passes.


\section{Conclusion and Future Work}\label{sec:conclusion}
In her doctoral thesis, Pitpimon Choorod presented a method which allows to distinguish Tor and non-Tor encrypted data traffic at high rates only on the basis of the analysis of hex digits occurring in a single encrypted data packet or their relative frequency. However, it is still not fully understood, why this is possible. One might think that this distinction is made possible by the fact that Tor traffic, unlike other encrypted traffic, is encrypted multiple times, but this would be in clear contradiction to the cryptographic theory of secure encryption. In this paper, we have deliberately omitted the technical network superstructure and concentrated solely on the distinction between single- and triple-encrypted data traffic, whereby we have also examined different operating modes for the \ac{AES} block cipher. The results are absolutely clear: with the proposed method none of the the three machine learning algorithms, Random Forest, Decision Tree, or k-Nearest Neighbor, is capable of distinguishing between single- and triple-encrypted data. These results are in accordance with crypto theory and illustrate that encryption is not the reason why a distinction can be made between Tor and non-Tor encrypted traffic.\par 
In order to better understand why this distinction is nevertheless possible, we will conduct further experiments in the future to gradually rule out possible explanations and identify the actual cause.

\renewcommand*{\bibfont}{\footnotesize}
\setlength{\labelnumberwidth}{0.45cm}
\printbibliography[notcategory=selfref]

\end{document}